\documentclass{ieeeaccess}
\usepackage{cite}
\usepackage{amsmath,amssymb,amsfonts}
\usepackage{algorithmic}
\usepackage{graphicx}
\usepackage{textcomp}
\usepackage{hhline}
\usepackage[caption=false,font=footnotesize]{subfig}

\usepackage{listings}
\usepackage{float}
\usepackage{rotating}
\usepackage{adjustbox}
\usepackage{amsmath,amssymb,amsfonts}
\usepackage{array}
\usepackage{cleveref}
\usepackage{textcase}
\usepackage{mathptmx}

\lstdefinestyle{ruleset}{
  basicstyle=\ttfamily\footnotesize,
  numbers=left,
  numberstyle=\tiny,
  frame=tb,
  framesep=2mm,  
  breaklines=true,
  breakatwhitespace=false,
  tabsize=2,
  columns=fullflexible,
  keepspaces=true,
  showstringspaces=false,
}

\def\BibTeX{{\rm B\kern-.05em{\sc i\kern-.025em b}\kern-.08em
    T\kern-.1667em\lower.7ex\hbox{E}\kern-.125emX}}

\makeatletter
\newdimen\xfigwd
\xfigwd=\columnwidth
\makeatother

\begin{document}
\history{}
\doi{}

\title{RuleSet Generation Framework for Application Layer Integration in Quantum Internet}
\author{
\uppercase{Rei Kawano}\authorrefmark{1},
\uppercase{Shin Nishio}\authorrefmark{1}\authorrefmark{2}, 
\uppercase{Hideaki Kawaguchi}\authorrefmark{1}, 
\uppercase{Shota Nagayama}\authorrefmark{3}, 
\uppercase{and Takahiko Satoh.}\authorrefmark{4}
}
\address[1]{Graduate School of Science and Technology, Keio University, Kanagawa, 223-8522, Japan}
\address[2]{Department of Physics \& Astronomy, University College London, London, WC1E 6BT, United Kingdom}
\address[3]{Graduate School of Media Design, Keio University, Yokohama, Kanagawa, 223-8526, Japan}
\address[4]{Faculty of Science and Technology, Keio University, Kanagawa, 223-8522, Japan}

\markboth
{Kawano \headeretal: RuleSet Generation Framework for Application Layer Integration in Quantum Internet}
{Kawano \headeretal: RuleSet Generation Framework for Application Layer Integration in Quantum Internet}

\corresp{Corresponding author: Rei Kawano (email: krr.kawano@keio.jp).}

\begin{abstract}
Layered architectures for the Quantum Internet have been proposed, inspired by that of the classical Internet, which has demonstrated high maintainability even in large-scale systems. While lower layers in the Quantum Internet, such as entanglement generation and distribution, have been extensively studied, the application layer — responsible for translating user requests into executable quantum-network operations — remains largely unexplored. A significant challenge is translating application-level requests into the concrete instructions executable at lower layers. In this work, we introduce a RuleSet-based framework that explicitly incorporates the application layer into the layered architecture of the Quantum Internet. Our framework builds on a RuleSet-based protocol, clarifying communication procedures, organizing application request information, and introducing new Rules for application execution by embedding application specifications into RuleSets. To evaluate feasibility, we constructed state machines from the generated RuleSets. This approach enables a transparent integration from the application layer down to the physical layer, thereby lowering barriers to deploying new applications on the Quantum Internet.
\end{abstract}


\titlepgskip=-15pt

\maketitle

\section{Introduction}
\PARstart{T}{\NoCaseChange{he}} Quantum Internet~\cite{kimble2008quantum}, built on the principles of quantum mechanics, is gaining attention as a new communication infrastructure that surpasses classical Internet systems. It holds the potential for diverse applications, including quantum key distribution (QKD)~\cite{bennett1984quantum}, distributed quantum computing, and blind quantum computing~\cite{broadbent2009universal}. In recent years, experimental testbeds for quantum entanglement distribution have been established internationally, advancing practical research in the field. These testbeds are no longer confined to laboratory-scale implementations but have expanded to full-scale network constructions in urban areas~\cite{kucera2024demonstration}~\cite{knaut2024entanglement}. These developments demonstrate that research on the physical layer and network foundations of the Quantum Internet is progressing steadily. In contrast, discussions surrounding the applications that operate on these foundational technologies remain at a much earlier stage.

\begin{figure}[htbp]
    \centering
    \begin{minipage}{0.49\linewidth}
        \centering
        \includegraphics[width=\linewidth]{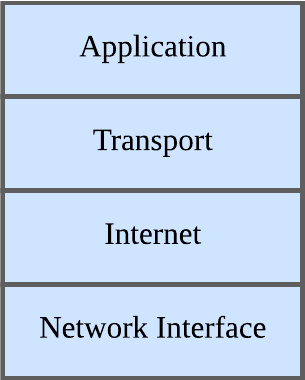}
        (a)
        \label{fig:clayer}
    \end{minipage}
    \hfill
    \begin{minipage}{0.49\linewidth}
        \centering
        \includegraphics[width=\linewidth]{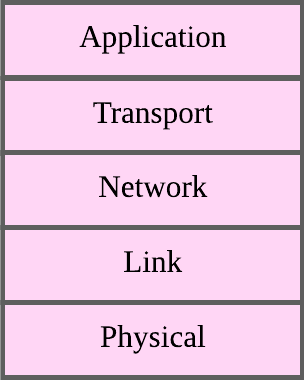}
        (b)
        \label{fig:qlayer}
    \end{minipage}
    \caption{The layered architectures for (a) classical (TCP/IP) and (b) Quantum Internet protocol.}
    \label{fig:layer}
\end{figure}

In the classical Internet, a hierarchical structure, as shown in Fig.~\ref{fig:layer}~(a)~\cite{fall2012tcp}, has been established as the network architecture, with protocols defined for each layer. Each resource operates according to these protocols, enabling connectivity between adjacent layers and allowing applications at the highest layer to be transparently connected to the hardware at the lowest layer, while ensuring high maintainability and extensibility

Similar to the classical model, a layered architecture has recently been proposed for the Quantum Internet. As one such proposal, the architecture shown in Fig.~\ref{fig:layer}~(b)~\cite{dahlberg2019link}~\cite{kozlowski2019towards} is introduced. The Physical Layer includes quantum devices and optical fibers responsible for generating, manipulating, and measuring qubits. The Link Layer manages entanglement sharing between adjacent nodes. The Network Layer handles long-distance entanglement using entanglement swapping~\cite{pan1998swap}. The Transport Layer is responsible for qubit transmission, and the Application Layer is intended to implement application protocols utilizing both classical and quantum networks~\cite{wehner2018quantum}. While alternative hierarchical structures have been suggested~\cite{van2008system, pirker2019quantum, Li2024}, their realization at the top-level application layer is still left for future work.

While specific application concepts have been proposed, their number is limited~\cite{zhang2024}, and clear benchmarks for the resources and cost metrics required for implementation have yet to be established. Furthermore, research on applications and hardware layers has progressed independently, resulting in a significant disconnect between the application layer and the lower layers. This disconnect highlights the lack of both a standardized methodology and a concrete approach for integrating applications with hardware.

In current quantum network architectures, protocol design has primarily been studied bottom-up, beginning with the lower layers, including the physical layer. This is largely because the overall performance of a quantum network is constrained by physical limitations~\cite{zhang2024}. Moreover, applications are frequently equated with the mere sharing of Bell pairs, under the assumption that once the physical layer is established, applications will function accordingly. However, significant challenges remain in executing nontrivial applications on a fully realized Quantum Internet.

Given this background, there is a clear need to integrate the application layer with lower layers in a transparent manner~\cite{Li2024}. In this work, we propose a framework that specifies the communication procedures and information transformations required to translate application-layer requests into instructions executable by lower layers of the Quantum Internet. Our goal is to establish the foundational architecture for this framework. 

In this work, we adopt a RuleSet-based protocol~\cite{matsuo2019quantum, satoh2023rula, van2022Internet}. NetQASM~\cite{dahlberg2022netqasm} has been proposed as a method for describing Quantum Internet applications and executing them between end nodes over the network. In NetQASM, the application layer decomposes quantum operations into subroutines, which are then executed sequentially on the quantum network processing unit. In contrast, the RuleSet-based protocol enables each node to autonomously proceed with operations according to predefined Rules. As a result, nodes no longer need to wait for decisions from the application layer, thereby minimizing latency more effectively.

\subsection{RuleSet-based protocol}
The RuleSet-based protocol is designed to synchronize network-wide operations with minimal classical communication. Since quantum entanglement serves as a critical resource in quantum communication, adjusting quantum operations at each node is essential to realize a quantum network. Such an adjustment requires classical communication. However, as communication distances increase, classical communication introduces significant delays, impacting the synchronization of operations. Therefore, to ensure scalability, the use of classical communication during quantum communication should be minimized.

By employing this protocol, each node can operate autonomously without waiting for instructions from other nodes, thereby reducing latency and enabling a high-performance network.

\begin{figure}[htbp]
\centerline{\includegraphics[width=\linewidth]{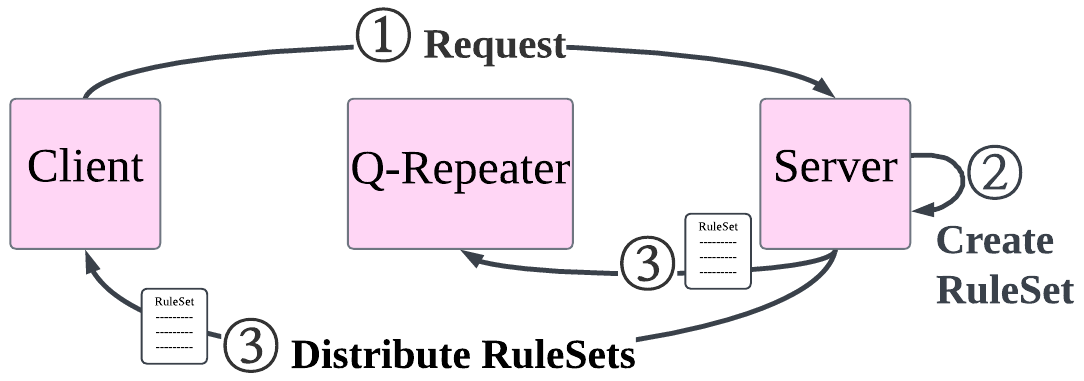}}
\caption{Classic communication procedure from application request to RuleSets distribution. \textcircled{1} Client sends application request to Server. \textcircled{2} Server creates RuleSets for all nodes involved in the connection. \textcircled{3} Server distributes RuleSets to other nodes. Q-Repeaters denote Quantum Repeaters, which collectively represent all repeaters on the connection between the client and server as a single node.}
\label{fig:overview}
\end{figure}

A RuleSet is an object held by all nodes in a connection and is used for making operational decisions. A RuleSet consists of one or more Rules, each of which contains a \textbf{Condition} and a corresponding \textbf{Action}. By performing all the operations specified in the Action, a single task is completed. RuleSets are created for each connection and distributed to all nodes involved in the connection before communication begins. By following the RuleSet they hold, each node contributes to achieving the desired overall operation of the network. The RuleSets for all nodes are generated by the end node acting as the server and then distributed (Fig.~\ref{fig:overview}).

\section{Layer Definition}
In this Section, we provide a clear definition of the application and transport layers within the quantum network stack and clarify the specific aspects addressed by our proposed framework. Zhang et al.~\cite{zhang2024} pointed out that the distinction between the transport layer and the application layer in quantum networks remains blurry. We also clarify the distinction between entanglement sharing across end nodes and actual application execution from the perspective of layer definitions. The specific roles of the application and transport layers are defined as follows.

\subsection{Transport Layer}
The transport layer is primarily responsible for quantum data transmission and entanglement management. Within the network stack, it is the highest layer that directly operates on qubits. 

The transport layer manages session control at the communication path level and handles fidelity management of quantum entanglement.

First, with regard to session management at the communication path level, similar to how the transport layer in the classical TCP/IP stack manages TCP sessions, the transport layer in the quantum network architecture is responsible for establishing, maintaining, and terminating communication paths.

Second, fidelity management of the entanglement between end nodes delivered from the network layer is performed. After the network layer generates entanglement with the required fidelity and passes it to the transport layer, the fidelity may degrade due to hardware-level decoherence when the entanglement is stored for a long time, resulting in entanglement that no longer satisfies the requested fidelity. Therefore, the entanglement handed over from the network layer is temporarily stored, and when it is requested, its fidelity is checked. If it no longer meets the requirement, it is replaced with entanglement that satisfies the requested fidelity. This functionality ensures that the application can always utilize entanglement with the required fidelity.

\subsection{Application Layer}
The application layer defines the protocols necessary to enable the execution of client-side applications on the Quantum Internet. It does not directly operate on qubits; instead, all interactions and management are carried out entirely on the classical plane. The application layer is responsible for the following three.

First, it is responsible for the exchange of application-related information. To request the execution of a desired application, the client communicates the necessary information to the server. 

Second, it is responsible for session management at the application level. For each application session, the communication counterpart and execution state are distinguished and maintained, while managing the necessary information and handling termination through timeouts or signals. In other words, application-specific information is consistently managed throughout the entire process, from initiation to termination.

Third, the RuleSet creation is responsible for managing the application execution schedule. This functionality is normally implemented at the server. The server generates an optimal RuleSet for all nodes along the communication path, taking into account the application request received from the client as well as path information such as the resource capacities of the repeaters and the performance of the links. Since the RuleSet contains complete information on what and when each node should execute, it defines the operational schedule of all nodes involved in the communication during application execution.

\subsection{Positioning of the Proposed Framework}
Current RuleSets include only the rules necessary for entanglement sharing between remote nodes~\cite{matsuo2019quantum}. Therefore, the RuleSet-based protocol can be understood as connecting the layers below the network layer in the quantum network architecture. Consequently, if application-level operations, including quantum teleportation, can be translated into RuleSets, it becomes possible to achieve transparent connectivity from the topmost application layer down to the physical layer. 

\begin{figure}[htbp]
\centerline{\includegraphics[width=\linewidth]{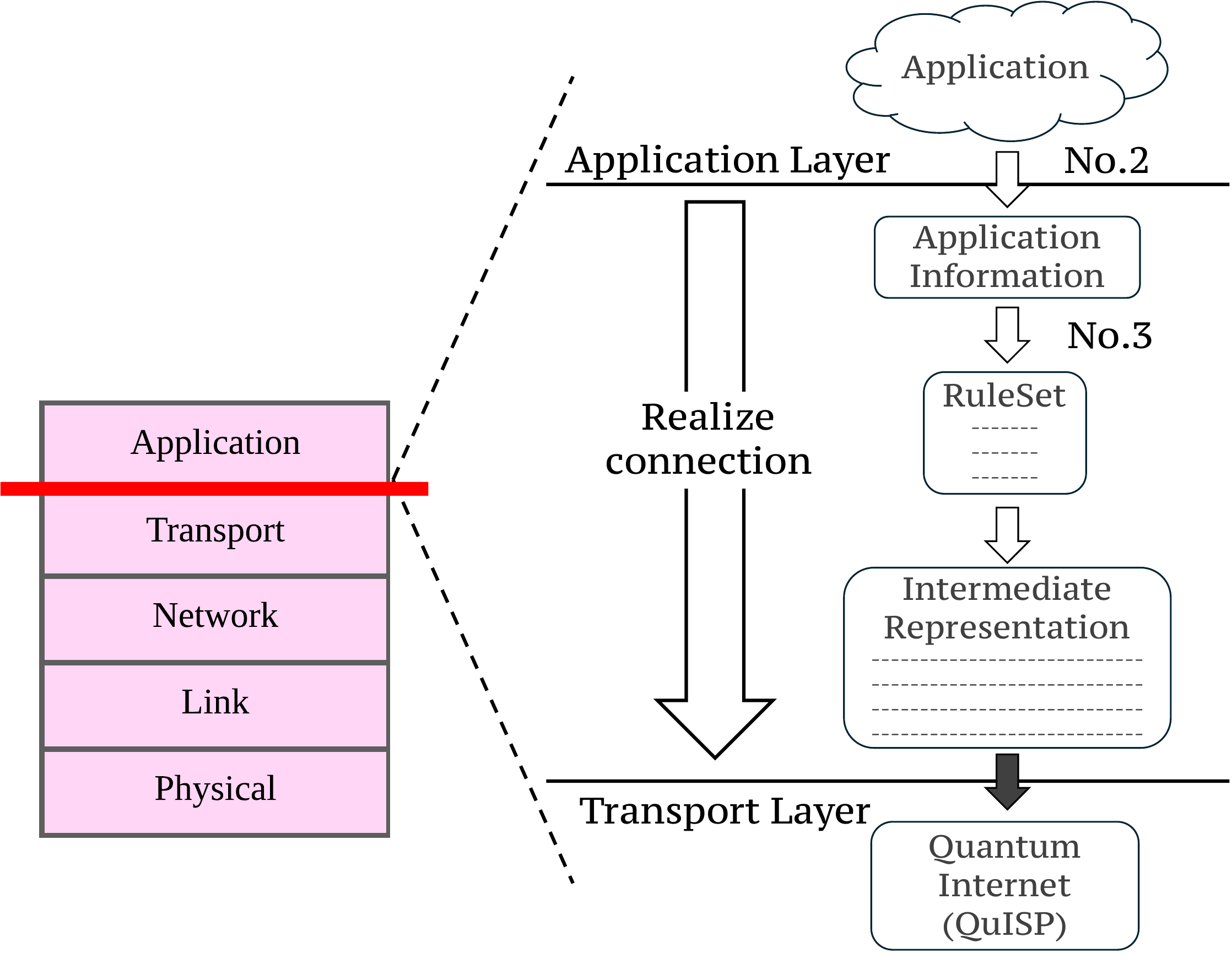}}
\caption{The process for connecting the Application layer and the Transport layer using the RuleSet-based protocol. By following this procedure, transparent integration between the application layer and the lower layers can be achieved. QuISP (Quantum Internet Simulation Package)~\cite{satoh2022quisp} is a large-scale quantum network simulator, which was adopted In this work as a virtual representation of the Quantum Internet.}
\label{fig:transaction}
\end{figure}

Based on the above, this study aims to clarify the following three key points: (i) the concrete communication procedure from the application request to the distribution of RuleSets to each node and the initiation of application execution, (ii) the identification of the minimal information required to construct the RuleSets, namely the application information that should be included in the application request, and (iii) the specific structure of the RuleSets generated based on the application information.

(i) In a client-server communication model, steps \textcircled{1} to \textcircled{3} in Fig.~\ref{fig:overview} illustrate how the RuleSet is distributed to all nodes along the communication path, enabling the application to enter an executable state~\cite{matsuo2019quantum}. However, the detailed interactions within these steps require further clarification. As shown in Fig.~\ref{fig:transaction}, the application undergoes information transformation during the procedure in Fig.~\ref{fig:overview}, eventually becoming a form acceptable to the lower layers. 

(ii) When executing an application, it is necessary to appropriately extract only the information essential for RuleSet generation from a large amount of available data. However, this required information is not self-evident. This corresponds to the transformation indicated by No.2 in Fig.~\ref{fig:transaction}.

(iii) Since existing RuleSets do not incorporate application-level context, the detailed description of RuleSets that include the execution phase of the application has not yet been explored. This corresponds to No.3 in Fig.~\ref{fig:transaction}.

These three issues will be addressed through proposals in the following section.
\section{Proposals}
\subsection{Problem setting}
We define the problem setting considered in this work. We focus on communication based on a client-server model. The client is assumed to possess only minimal quantum operation capabilities, while the server is assumed to be a node capable of performing quantum computation, such as fault-tolerant quantum computing (FTQC). In other words, we envision a scenario in which the client remotely accesses a quantum computer. Although, in practice, the client and server would be connected via multiple quantum repeaters~\cite{briegel1998quantum} within the broader Quantum Internet, we simplify the model by treating these intermediate repeaters as a single entity, denoted as Q-Repeaters. As such, we consider communication among three nodes: the client, the Q-Repeaters, and the server. Furthermore, we assume an idealized Quantum Internet in which all nodes are equipped with quantum memory~\cite{chou2007functional, sangouard2011quantum}.

\subsection{Procedure for Application Preparation}
\begin{figure}[htbp]
\centerline{\includegraphics[width=\linewidth]{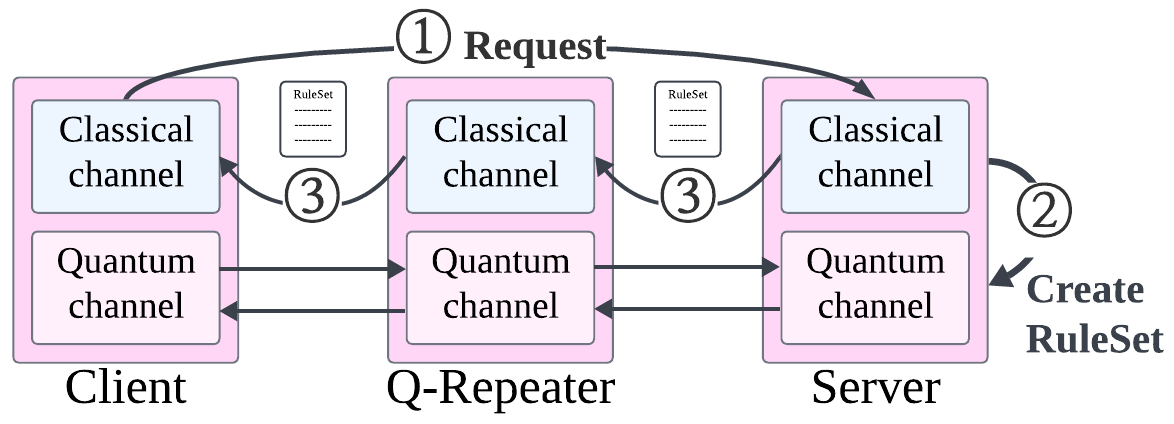}}
\caption{Adaptation of the nodes in Fig.~\ref{fig:overview} to the structure of actual Quantum Internet nodes. Each node has a classical channel and a quantum channel.}
\label{fig:cqoverview}
\end{figure}

\begin{figure*}[htbp]
\centerline{\includegraphics[width=\linewidth]{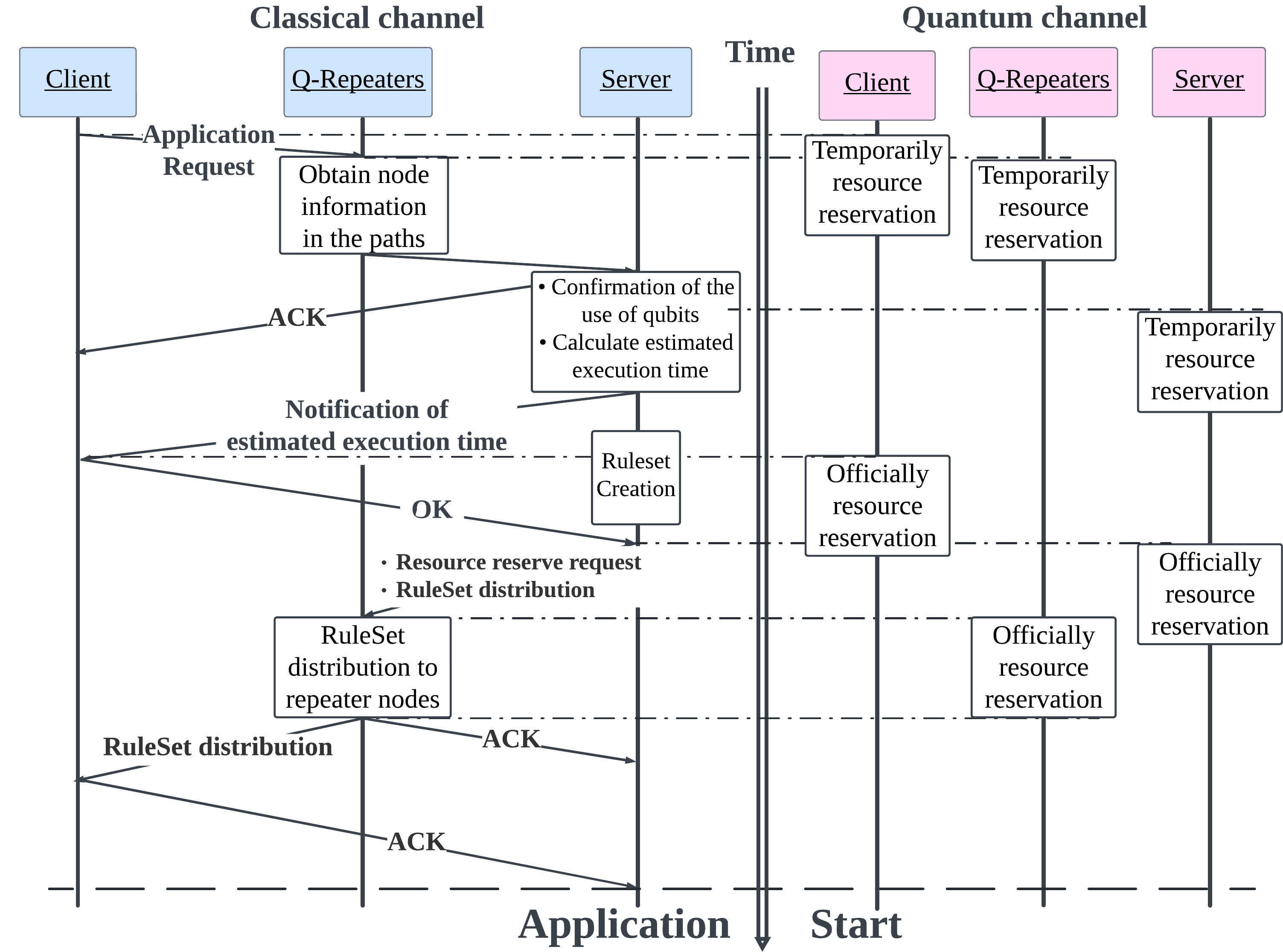}}
\caption{Communication Procedure for Application Preparation. This figure shows the detailed communication procedures between the client, server, and the Quantum Repeaters (Q-Repeaters) until the application preparation is completed.}
%for executing the quantum application is completed.}
\label{fig:prep}
\end{figure*}

This section examines the specific communication procedures involved from the application request to the distribution of the RuleSet across nodes and the start of application execution. The process follows the steps illustrated in Fig.~\ref{fig:overview}~\cite{matsuo2019quantum}, but the detailed interactions within these steps remain unclear. To address this, Fig.~\ref{fig:cqoverview} presents a revised version of Fig.~\ref{fig:overview} that better reflects actual nodes in a Quantum Internet environment. Each node is equipped with a classical channel for classical communication and a quantum channel for quantum communication, as classical communication is essential for quantum communication. In this way, Quantum Internet applications differ significantly from classical applications in that they require not only quantum communication but also hybrid of quantum and classical communication. This hybrid nature is expected to make the scheduling of application execution more complex than that of classical applications. To clarify this, Fig.~\ref{fig:prep} illustrates the actual interactions exchanged among the three nodes.

In clarifying these interactions, the necessary information exchanges before application execution were considered. First, the information required by the server to create the RuleSet must be securely exchanged. Second, the client must be informed of the estimated execution time of the application, as assessed by the server, prior to execution, enabling the client to decide whether to proceed with the application. Additionally, it is anticipated that the early-stage Quantum Internet will lack sufficient quantum resources, including quantum memory and qubits for quantum computing. To stabilize application execution, a resource reservation approach may be adopted, where each application secures the necessary resources before execution, and these resources are not used by other applications during execution. This approach is conceptually similar to circuit-switching~\cite{leon2004communication} in classical communication systems.

In Fig.~\ref{fig:prep}, the left half represents the classical communication channel, while the right half represents the quantum communication channel. To clarify the interactions of classical and quantum communication, Fig.~\ref{fig:prep} separates the classical and quantum channels. The client (or Q-Repeaters, Server) on the classical channel side and the client (or Q-Repeaters, Server) on the quantum channel side in Fig.~\ref{fig:prep} correspond to the same node, as shown in Fig.~\ref{fig:cqoverview}. Additionally, the vertical axis represents time, and the timelines for classical and quantum communications are synchronized.

\setlength{\labelwidth}{4cm}
\subsection{Functions over the Classical channel}
\subsubsection{Obtain node information in the paths} While transmitting the application request to the server, information about the path it traverses is collected. This path information includes the performance of the optical fibers between nodes and the available quantum memory at each node.

\subsubsection{Confirmation of the use of qubits}
The server checks the usage status of qubits within the quantum computer and identifies the available qubits and quantum memory.

\subsubsection{Calculate estimated execution time}
The server calculates the estimated execution time for the requested application based on the acquired path information and the availability of quantum computing resources. This estimated execution time is derived by considering factors such as the entanglement supply rate at the required fidelity, the duration of quantum operations, classical communication delays, and classical signal processing time. 

\subsubsection{Notification of the estimated execution time} The server sends the estimated execution time of the application to the client. The client uses this information to decide whether to execute the application and notifies the server of its decision. If the application is to be executed, the process moves on to the subsequent steps. If not, the connection is terminated, and the request information as well as all RuleSets created during the "Creation of RuleSet" step are discarded.

\subsection{Functions over the Quantum Channel}    
\subsubsection{Temporarily resource reservation} Nodes along the path temporarily reserve resources for use by the application. On the client node, in anticipation of executing multiple applications, memory allocation for each application is configured and temporarily reserved at the time the request is sent. When a repeater node receives an application request from the client, it checks the maximum amount of memory available for the application at that moment, appends this information to the request, and forwards it to the next node. During this process, the memory marked as available is temporarily reserved to prepare for application execution. Similarly, at the server, the resources marked as available during the "Confirmation of the use of qubits" step are also temporarily reserved. This mechanism ensures that the required resources are securely reserved for application execution.

If a temporary reservation mechanism is not implemented, it would be necessary to re-query the Q-Repeaters for available resources before RuleSet generation, which would complicate the procedure. If resources are consumed by another application before the execution decision, the application may no longer meet its estimated execution time. Practical problems that may result from this function, along with possible countermeasures, will be described in a subsequent section.

\subsubsection{Officially resource reservation} Upon receiving the application execution command from the client, resources are officially reserved. At the client and repeater nodes, only the portion of the temporarily reserved memory specified by the server is officially reserved, while the remaining memory is released. Similarly, at the server, only the resources needed for the application are officially reserved, and any unnecessary resources are released.

This proposal focuses on clarifying the procedures necessary for application preparation among the client, server, and Q-Repeaters. Therefore, it does not consider time management in detail, such as time synchronization between classical and quantum channels. In practice, strict time management will be required when executing the application.

\subsection{Information Required for RuleSet Generation}
This section organizes the information required for the server to generate a RuleSet. It is crucial that the information sent from the client to the server is both necessary and sufficient for the server to generate the RuleSet. If the information is insufficient, the server will be unable to create a RuleSet that aligns with the application desired by the client. On the other hand, if the information is excessive, unnecessary details about the client could be disclosed to the server, which is undesirable when using the server remotely.

Table~\ref{tab:info_list} summarizes the required information, with illustrative application examples.

\begin{table}[ht]
\caption{A proposed list of necessary and sufficient information for application requests and their examples.}
    \centering
    \begin{tabular}{|c|c|c|}
        \hline
        Information & \multicolumn{2}{c|}{Examples}  \\ 
        \hhline{|=|=|=|} 
        Server Address & qc001 & qc002 \\ \hline
        Application Type & T(Blind-VQE) & T(Teleportation) \\ \hline
        Data Type & Bell pair & Bell pair \\ \hline
        Fidelity & 0.8 & 0.95 \\ \hline
        Resource / shot & 10,000 & 1 \\ \hline
        \# of shot & 4,000,000 & 100 \\ \hline
        Est. execution time & 6 hours & 1 minutes\\ \hline
    \end{tabular}
    \label{tab:info_list}
\end{table}

\begin{table*}[htbp] 
\caption{List of application types. The 'W' label indicates that the process must wait until certain conditions are met before proceeding to the next step, while the 'I' label signifies that the process can immediately proceed to the next step without waiting. (Reproduced from~\cite{van2017optimizing})}
    \centering
    \begin{tabular}{|c|l||c|c|}
        \hline
         Class & Name & Reception success notification & Pauli frame propagation \\ \hline
         B & Bell (Bell inequality violations, QKD) & I & I  \\ \hline
         C & Clifford group computation & W & I \\ \hline
         T & General computation &  W & W \\
          & (including non-Clifford operations), teleportation &  &  \\ \hline
    \end{tabular}
    \label{tab:apptype}
\end{table*}

Each item is described as follows. 'Server Address' specifies the address of the server that the client accesses remotely. It is assumed that the client will access a server equipped with hardware suitable for the specific application. 'Application Type' is a categorized class that specifies how the application intends to utilize the network. The application types are shown in Table~\ref{tab:apptype}. Application types can be categorized based on the timing of Pauli frame corrections for entanglement-based measurement operations~\cite{van2017optimizing}.The blind variational quantum eigensolver (VQE) and teleportation listed in Table~\ref{tab:info_list} are class T, which involve non-Clifford operations requiring all Pauli corrections to be performed before measurement. In addition, applications that allow post-selection are categorized as class B, and those based on Clifford operations are categorized as class C. The application type has a significant influence on the scheduling of application execution at the end nodes.

'Data Type' refers to the type of quantum resource required, such as Bell pairs or GHZ states. 'Fidelity' represents the minimum fidelity acceptable for the application. Since fidelity and resource generation time are in a trade-off relationship, the requirements are expected to vary depending on the nature of the application.

'Resource/shot' and '\# of shot' indicate the number of resources required. A "shot" refers to one iteration of the process in VQE, namely “quantum circuit execution → classical learning → feedback to the next execution.” Since entanglement generation is time-consuming, it is more efficient to generate resources in parallel with the computational part of the application. Therefore, each shot is executed as soon as the resources required for that shot are generated, while resources for the next shot are prepared concurrently. As a result, the total number of required resources is given by $(\text{Resource/shot}) \times (\text{\# of shot})$.

'Est. execution time' refers to the application completion time requested by the client. To be viable as an application, it is important that execution is completed within the time allowed by the user, not whether it can theoretically be executed in real time.

\subsection{Structure of RuleSet}
This section examines how a RuleSet is specifically described based on the information contained in an application request. While RuleSets for sharing Bell pairs between end nodes already exist, the detailed description of RuleSets that also include application execution has not been fully explored. Therefore, we investigated how RuleSets incorporating application request information would be described.

\begin{figure}[htbp]
    \centering
    \includegraphics[width=\linewidth]{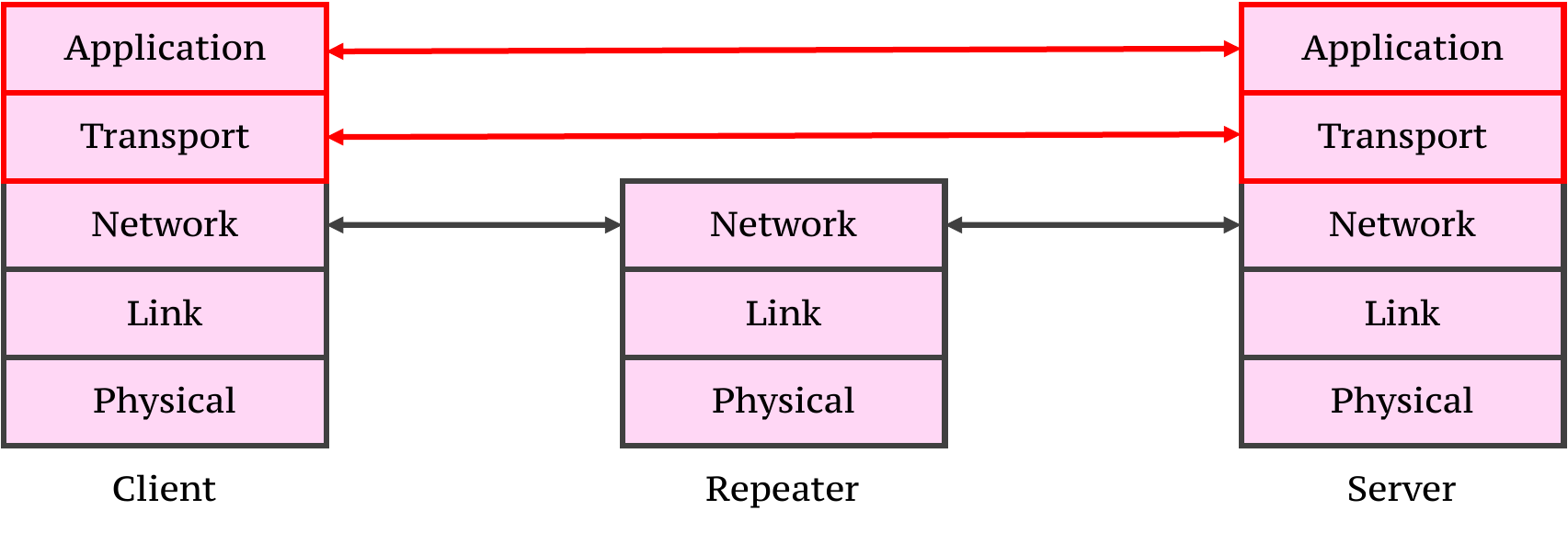}
    \caption{The protocol stack at each node in quantum communication and its interactions.}
    \label{fig:layer_interaction}
\end{figure}

The RuleSet follows a top-down approach, where the Rules within the RuleSet are executed sequentially from top to bottom~\cite{matsuo2019quantum}. Furthermore, the RuleSet includes termination conditions, allowing it to function as a counter for discarding the RuleSet and closing the connection. All RuleSets for the same connection share a common identifier. This identifier is also included in classical messages exchanged during the connection. As a result, even when different applications are running on various nodes, classical messages can still be correctly associated with their respective connections.

\begin{figure}[htbp]
\centerline{\includegraphics[width=\linewidth]{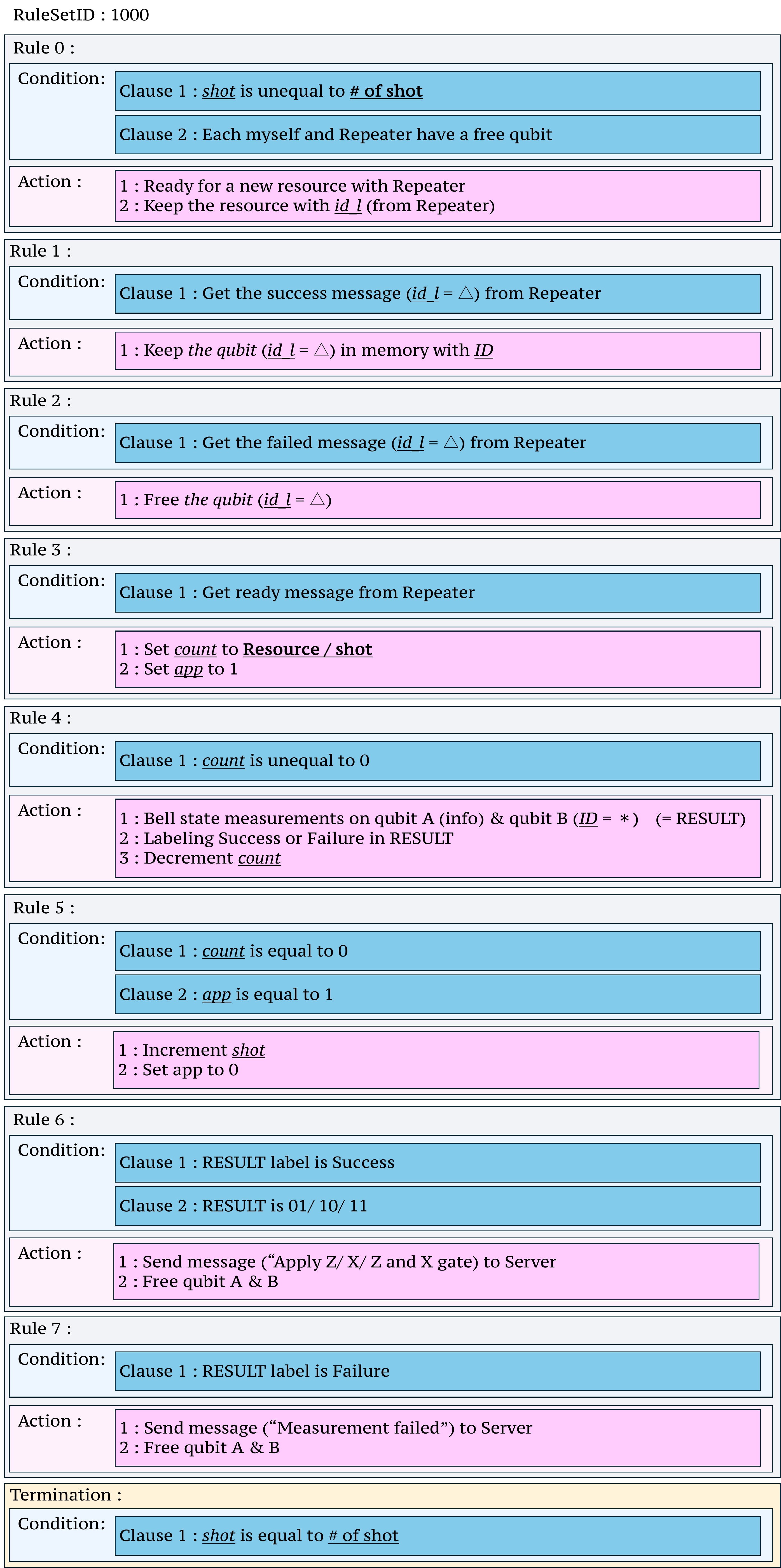}}
\caption{Client RuleSet for quantum teleportation. Rule 0 through Rule 3 handle entanglement sharing with the Repeater and perform operations based on the outcomes of entanglement swapping. Rule 4 through Rule 7 involve measurement operations for teleportation, used to transfer quantum information to the server.}
\label{fig:clientruleset}
\end{figure}

Based on the above, Fig.~\ref{fig:clientruleset} shows the Client RuleSet for executing quantum teleportation between the client and the server in a configuration consisting of three nodes: client, repeater, and server. Details of the variables defined within this RuleSet are provided in Table~\ref{tab:cli_variable}.

We begin by explaining the overall flow across all nodes. Fig.~\ref{fig:layer_interaction} illustrates the network architecture at each node. Based on the layer definitions, the transport and application layers are implemented only at the end nodes. Since existing RuleSets are limited to descriptions up to the network layer, the new Rules proposed pertain to the context at or above the transport layer at the end nodes. 

The variables `count` and `shot` are used to manage the number of consumed resources and completed shots, respectively.  The variable `app` serves as a flag indicating whether a shot is currently being executed. The variables `id\_l` and `ID` are entanglement identifiers, where `id\_l` is shared with an adjacent node, and `ID` is shared with a newly connected node after entanglement swapping.

\begin{table}[htbp]
    \centering
    \caption{Variables included in the RuleSet in Fig.~\ref{fig:clientruleset}.}
    \begin{tabular}{|c|c|}
        \hline
         Variables & Example value \\ \hline
         shot & $0 \sim$ \# of shot \\ \hline
         count &  $0 \sim$ Resource/shot \\ \hline
         app & 0 / 1 \\  \hline
         id\_l &  1001\\ \hline
         ID &  2001\\ \hline
    \end{tabular}
    \label{tab:cli_variable}
\end{table}

\subsection{Intermediate Representation (IR) conversion}
This section examines how the RuleSets distributed to each node are executed on the hardware of the respective nodes. Here, we consider QuISP~\cite{satoh2022quisp, matsuo2019simulationdynamicrulesetbasedquantum}, a large-scale quantum network simulator that adopts a RuleSet-based protocol, as a representative model of a Quantum Internet. QuISP is an event-driven simulator for quantum repeater networks and is regarded as a foundational platform for the next-generation Quantum Internet. In QuISP, a RuleSet is transformed into an Intermediate Representation (IR) at each node, enabling execution on the underlying hardware. The IR functions as a simple labeled assembly-like language, allowing the specification of more fine-grained hardware-level operations.

The current implementation of QuISP focuses primarily on lower-layer protocol design, and thus, it does not yet support application-layer functionality that utilizes entanglement after its generation between end nodes. Consequently, IR corresponding to application contexts has not been implemented. In this work, we define several new IRs to enable the transformation of RuleSets containing application context into executable forms. 

These IRs are expected to be supported in future versions of QuISP. The newly defined IRs are presented Fig.\cref{lst:send_request,lst:wait,lst:qubit_pair,lst:generate_entangle,lst:store,lst:notify}, while existing IRs are omitted.

\begin{figure}[htbp]
\begin{lstlisting}[style=ruleset, linewidth=\linewidth]
SEND_REQUEST <partner_address> <request>
// example
SEND_REQUEST partner_addr "FREE_QUBIT_REQUEST"
\end{lstlisting}
\caption{Send request to partner node.}
\label{lst:send_request}
\end{figure}

\begin{figure}[htbp]
\begin{lstlisting}[style=ruleset, linewidth=\linewidth]
WAIT_FOR_RESPONSE <partner_address> <qubit_register>
// example
WAIT_FOR_RESPONSE partner_addr qubit_id_partner
\end{lstlisting}
\caption{Await partner's response.}
\label{lst:wait}
\end{figure}

\begin{figure}[htbp]
\begin{lstlisting}[style=ruleset, linewidth=\linewidth]
CREATE_PAIR <structure_name> <qubit1_id> <qubit2_id>
// example
CREATE_PAIR pair qubit_id_self qubit_id_partner
\end{lstlisting}
\caption{Create a qubit\_ID pair structure for entanglement swapping.}
\label{lst:qubit_pair}
\end{figure}

\begin{figure}[htbp]
\begin{lstlisting}[style=ruleset, linewidth=\linewidth]
GENERATE_ENTANGLEMENT <qubit_id_self> <qubit_id_partner> <fidelity>
// example
GENERATE_ENTANGLEMENT qubit_self qubit_partner 0.95
\end{lstlisting}
\caption{Generate entanglement between nodes.}
\label{lst:generate_entangle}
\end{figure}

\begin{figure}[htbp]
\begin{lstlisting}[style=ruleset, linewidth=\linewidth]
REGISTER_ENTANGLEMENT <entanglement_id> <qubit_id_self> <qubit_id_partner> <fidelity>
// example
REGISTER_ENTANGLEMENT id_r qubit_self qubit_partner 0.95
\end{lstlisting}
\caption{Store generated entanglement to memory.}
\label{lst:store}
\end{figure}

\begin{figure}[htbp]
\begin{lstlisting}[style=ruleset, linewidth=\linewidth]
SEND_READY <left_partner_address> <right_partner_address>
// example
SEND_READY left_partner right_partner
\end{lstlisting}
\caption{Notify both ends of resource readiness.}
\label{lst:notify}
\end{figure}

Through the newly defined IRs, the following functionalities have been added: the ability to generate entanglement that meets the fidelity requirements specified by the application, the ability to store entangled states in memory, and the ability to separate operations into distinct stages—distinguishing between entanglement generation and application execution. 

By incorporating them, the proposed RuleSet is transformed as shown in Fig.~\ref{lst:IR}.

\begin{figure}[htbp]
\begin{lstlisting}[style=ruleset, linewidth=\linewidth]
Condition
  LOAD count "count"
  BNQ count 0 PASSED
  RET COND_FAILED

PASSED:
  RET COND_PASSED

Action
qubit: q0, q1
reg: pauli_op, result
key: "sent_swap_message_{shared_rule}"

START:
  SET pauli_op 0
  LOAD count "count"
  GET_QUBIT_BY_SEQ_NO q0 myself id_data
  GET_QUBIT_BY_SEQ_NO q1 new_partner_addr ID
  GATE_CNOT q0 q1
  GATE_H q1
  MEASURE pauli_op 0 q0 x
  MEASURE pauli_op 1 q1 z
  FREE_QUBIT q0
  FREE_QUBIT q1
  GET_RESULT q0 q1
  BRANCH_IF_SUCCESS SUCCESS
  JMP FAIL

SUCCESS:
  SEND_RESULT new_partner_addr "SUCCESS" pauli_op ID
  INC count
  STORE "count" count
  RET FINISHED

FAIL:
  SEND_SWAPPING_RESULT new_partner_addr "FAIL" ID
  RET FINISHED
\end{lstlisting}
\caption{Rule 4/6/7: Perform Bell measurement.}
\label{lst:IR}
\end{figure}

\section{Evaluation}
In this section, we evaluate the effectiveness of the proposed method from two perspectives. First, we examine the validity of the generated IR by verifying whether it correctly represents the intended processing procedures. Second, we assess the executability of the RuleSet and the validity of the estimated execution time through comparison with simulation results.

\subsection{Validity of IRs}
\begin{figure}
    \centering
    \includegraphics[width=\linewidth]{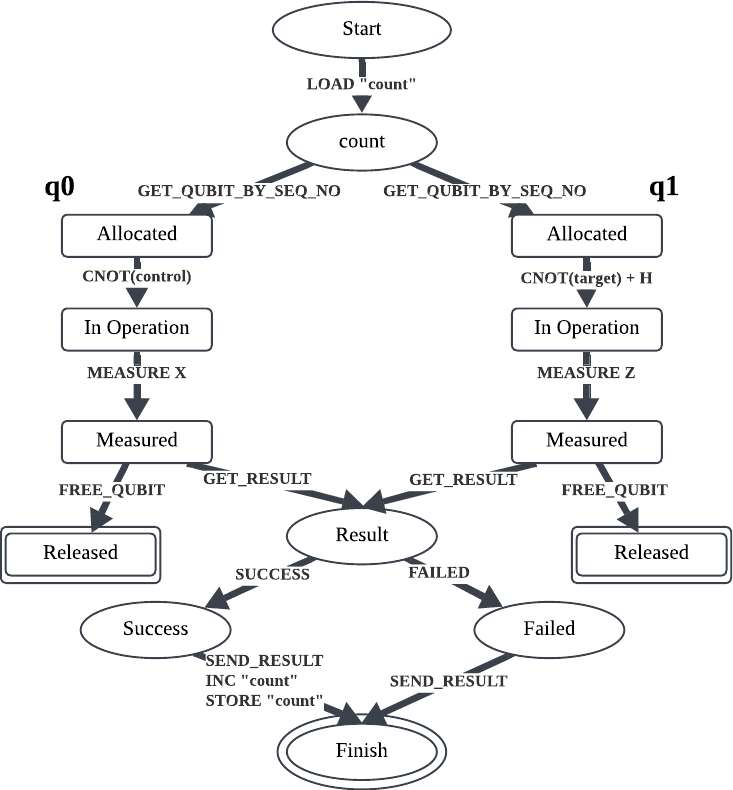}
    \caption{State machine based on the IR in Listing~\ref{lst:IR}. Squares represent quantum operations, and circles represent classical operations.}
    \label{fig:state_machine}
\end{figure}

The validity of the IR generated based on the proposed method was verified using a state machine. Once a RuleSet is converted into IR, a state machine can be constructed from the IR. This allows us to verify whether the intended behavior is correctly performed, and whether qubits are properly released or updated to the next Rule, thereby enabling a static evaluation of the generated RuleSet. Fig.~\ref{fig:state_machine} shows the state machine constructed from the IR in Listing~\ref{lst:IR}. In Fig.~\ref{fig:state_machine}, we confirmed that all qubits are released after measurement, that both classical and quantum processes reach their termination states, and that all conditional branches are comprehensively described. Through this verification, the validity of the RuleSets generated by the proposed method can be qualitatively validated.

We next examine the extensibility of the IR. In this work, we only focused on quantum teleportation; however, there exist various other Quantum Internet applications, such as blind quantum computation, QKD, distributed quantum computation, and quantum sensing. To execute these applications, the IR required at the end nodes includes local gate operations, measurement operations, resource request functionality, classical communication functionality such as transmitting measurement results, memory management functionality, and local computation functionality (classical computation and quantum circuit execution) that is completed within a node. Among these, the resource request and memory management functionalities can be supported by the newly defined IRs, while the others can be expressed using the existing IRs. Therefore, applications other than teleportation can also be supported by the newly defined IRs, and the IR proposed in this work possesses sufficient extensibility for Quantum Internet applications in general.

\subsection{Simulation Results}
\begin{figure}
    \centering
    \includegraphics[width=\linewidth]{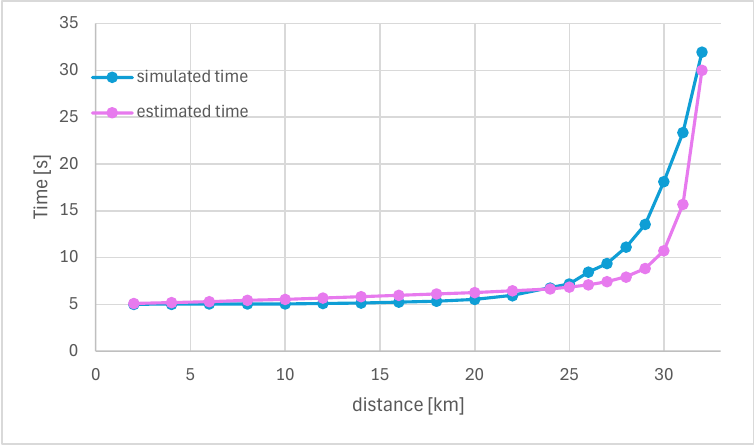}
    \caption{Comparison of the execution time estimated by the server and the execution time obtained with the QuISP simulator. Distance indicates the total link length between the end nodes.}
    \label{fig:sim_result}
\end{figure}

Next, Fig.~\ref{fig:sim_result} shows a comparison between the execution time as estimated by the server and the execution time obtained with the QuISP simulator using the RuleSet for quantum teleportation generated by the proposed protocol. In the current version of QuISP, the functionality to accept applications that use entanglement after its generation between end nodes has not yet been implemented; thus, the execution is limited to entanglement generation. In the case of quantum teleportation, the time required for the teleportation operations is much shorter than that for Bell pair generation, and therefore, Bell pair generation is the dominant factor in the execution time.

The simulation was conducted with the fiber loss set to 0.2 dB/km, the coupling efficiency and the detection efficiency set to 0.8. The number of memories at the end nodes was set to 10, and that at the repeater was set to 7. The total number of resources was set to 1000. The classical communication delay was assumed based on the speed of light in the medium, and the initial setup time was set to 5 s. For the node distances, a Bell-state analyzer (BSA) was placed at the midpoint between the end node and the repeater, and the link lengths on both sides of the repeater were set to be equal. All other parameters followed the default settings of QuISP.

As shown in Fig.~\ref{fig:sim_result}, the simulation time is well estimated for end-to-end distances up to $32$ km, indicating the validity of the estimated execution time calculation. Furthermore, for distances up to approximately $24$–$25$ km, a simple linear loss model was sufficient for estimation; beyond this range, however, it became necessary to account for the exponential degradation in success probability caused by fiber loss and jitter fluctuations. In particular, as the link distance increases, variations in photon arrival times due to jitter and dispersion become more pronounced, reducing the probability of detecting both photons simultaneously within the coincidence window, and thus lowering the success probability compared to that predicted by a simple loss model. Nevertheless, overly strict estimation of execution time is undesirable, as the computation itself would require excessive time. Therefore, it is expected that the estimation should be performed using an appropriate model selected according to the link distance.

\section{Toward Practical Use: \\ Issues and Countermeasures}
In this section, we address the issues that may arise in the practical use of the proposed protocol and discuss possible countermeasures. The countermeasures presented here are tentative, and the optimal solutions remain to be examined in future work.

\subsection{Management of Multiple Application RuleSets in a Node}
First, we consider the management of cases where a single node holds multiple RuleSets. Although a FIFO policy, in which RuleSets are processed in the order of arrival, could be adopted, here we focus on scheduling based on priorities. While FIFO is simple to implement, it has limitations: if a RuleSet requires a long execution time, other executions may be blocked; moreover, if the condition of the next Rule to be executed is not satisfied, the overall throughput may decrease. Therefore, priority-based scheduling allows for more flexible and efficient execution. The server assigns quantitative priority labels to RuleSets based on the requested execution time at the time of RuleSet generation, and each node queues the RuleSets accordingly.

\subsection{Handling Execution Beyond the Scheduled Time}
Finally, we address the handling of cases where execution does not complete within the scheduled time. In our proposal, the expected execution time is pre-estimated, and the nodes involved in the connection assume that execution will be completed within this scheduled period. Therefore, if execution does not finish on time, some countermeasures are required.

To this end, we define two application types: Time-Bounded Execution (TBE) and Repeat Until Success (RUS). In the TBE type, all nodes terminate execution once the scheduled time has elapsed, and the client is informed of the intermediate results obtained up to that point. In the RUS type, execution continues beyond the scheduled time until all tasks included in the request are completed. Information regarding the application type can be included in the initial application request.

\subsection{Resource Contention Issues in Resource Reservation}
Next, we address issues related to resource reservation, focusing primarily on two cases: (\romannumeral 1) the problem of specific applications monopolizing all resources of a node, and (\romannumeral 2) the problem of inappropriate resource blocking due to the tentative reservation mechanism.

(\romannumeral 1) This issue arises in two cases: when a single application monopolizes all resources on a node, and when multiple long-running RUS-type applications are allocated in a way that effectively leaves the node in a monopolized state. To address the first case, we partition each node's memory into TBE and RUS regions. This separation prevents strictly managed TBE applications from becoming unexecutable due to resource exhaustion caused by RUS applications. And for the second case, we propose setting an upper bound on the amount of memory that a repeater can allocate to a single application. Since the resource capacity differs among repeaters, this upper bound is defined not as an absolute value but as a proportion of each node’s maximum memory capacity.

(\romannumeral 2) In our proposal, resources are tentatively reserved before an official reservation is applied. If this tentative state is not properly managed, unused resources may remain unavailable. To mitigate this issue, we introduce a timeout mechanism for tentative reservation. The timeout duration is determined by taking into account the calculation time of the estimated execution time, the time required for the client’s execution decision, as well as the classical communication delay occurring during this period.

\section{Discussion}
In the layer definition of the transport layer, similar to TCP in classical network protocols, the fidelity of entanglement is guaranteed so as to satisfy application requirements. This approach, however, implies discarding entanglement that has undergone decoherence~\cite{li2021efficient}, which is undesirable from the perspective of resource efficiency. Therefore, it is also worth considering the coexistence of a protocol that does not guarantee entanglement quality, analogous to UDP in classical network protocols. In such a case, when the required fidelity is no longer satisfied, the application layer could, for example, perform distillation at the application level.

In this work, the protocol design assumed an ideal Quantum Internet where all nodes, including clients, servers, and repeaters, are equipped with a sufficient amount of quantum memory. However, given the current state of technological development, it will likely take considerable time before quantum memory becomes available at all nodes~\cite{li2019nomemory}. Moreover, the presence or absence of quantum memory significantly affects factors such as time constraints, necessitating substantial modifications to protocol design. Looking ahead, designing protocols for applications on earlier generations of Quantum Internet, where quantum memory is not yet fully implemented, will serve as a practical guideline for running applications in the near future. Such efforts could bridge the gap between the current technological limitations and the eventual realization of a fully functional Quantum Internet.

Furthermore, a resource reservation method was introduced in this work to provide stable application execution. However, if the number of clients, servers, and intermediate nodes increases in the future, a complete resource reservation approach may lack scalability and might not be preferable. Recently, best-effort approaches inspired by the packet-switching model of the classical Internet have also been proposed~\cite{Bacciottini_2025}. Therefore, how to secure resources when the number of nodes increases, as well as whether resource reservation should be employed, remain issues to be discussed in future work.

\section{Conclusion}
In this work, we proposed a RuleSet generation framework to unify Quantum Internet applications with lower layers. Specifically, we clarified the communication procedures required before executing applications on the Quantum Internet, organized the information required for the lower layer to accept applications, proposed a method for generating RuleSets based on application requirements, and defined new IRs for the application context. The generated RuleSets were verified through state machines, and the executability was evaluated by simulations in QuISP. The contributions of this research enhance the practicality and consistency of connecting the application layer with lower layers and provide a foundational framework for future research and development in both the application and hardware layers of the Quantum Internet.

\section*{Acknowledgment}
This work was supported by JST Moonshot R\&D Grant Number JPMJMS226C.
Shin Nishio and TS were supported by the New Energy and Industrial Technology Development Organization (NEDO).
Shin Nishio was also supported by JSPS Overseas Research Fellowship.
TS was also supported by JMEXT KAKENHI Grant Number 22K1978.

\bibliographystyle{IEEEtran}
\bibliography{kawano}
\EOD

\end{document}